\documentclass[prl,aps,showpacs]{revtex4}

\usepackage{graphicx}
\usepackage{latexsym}
\usepackage{amsmath,amscd}
\usepackage{times}

\begin{document}

\title{Optimal Time to Sell a Stock in Black-Scholes Model: Comment on ``Thou shall buy and hold'', by A. 
Shiryaev, Z. Xu and X.Y. Zhou}
\author{Satya N. Majumdar$^{1}$ and Jean-Philippe Bouchaud$^{2}$}

\affiliation{
$^1$ Laboratoire de Physique Th\'eorique et
Mod\`eles Statistiques (UMR 8626 du CNRS),
Universit\'e Paris-Sud, B\^at. 100, 91405 Orsay Cedex, France\\
$^2$  Science \& Finance, Capital Fund Management, 6 Bd
Haussmann, 75009 Paris, France.}

\begin{abstract}
We reconsider the problem of optimal time to sell a stock studied by Shiryaev, Xu and Zhou \cite{SXZ}
using path integral methods. This method allows us to confirm the results obtained by these
authors and extend them to a parameter region inaccessible to the method used in  \cite{SXZ}. We also obtain the full 
distribution of the time $t_m$ at which the maximum of the price is reached for arbitrary values of
the drift.

\end{abstract}

\maketitle
\date{\today}

\section{Introduction}

In the preceeding paper, A. Shiryaev, Z. Xu and X.Y. Zhou \cite{SXZ} ask about the optimal time to sell a 
stock
over a certain time interval $[0,T]$, knowing that the price is a geometrical Brownian motion with 
a certain average return over the risk-free rate $a-r$ and a certain volatility $\sigma$. The answer 
to this question depends on the value of the adimensional parameter $\alpha=(a-r)/\sigma^2$. The method
used by the authors allow them to prove that whenever $\alpha > 1/2$ the optimal selling time $\tau^*$ 
is always at the end of the interval, $\tau^*=T$, whereas $\tau^*=0$ in the case $\alpha < 0$. In 
financial words, ``good'' stocks with a sufficiently large average return should be sold as late as
possible, whereas one should immediately get rid of ``bad stocks''. These results are clearly very
interesting; however one feels unsatisfied by the fact that the authors' method do not allow them
to treat the case $0 < \alpha \leq 1/2$. They discuss this point in the conclusion, mentioning (a) 
a working paper \cite{SXZ2} based on an alternative method showing that one should in fact sell immediately
as soon as $\alpha < 1/2$ and (b) that the case $\alpha < 1/2$ is not interesting financially because
``most stocks realize $\alpha > 1/2$ by a large margin''. 

The aim of this short note is to reconsider the problem using path integral methods which are well known in
physics but perhaps less well known in financial mathematics. This method allows one to treat all values
of $\alpha$ on the same footing. We confirm the results of \cite{SXZ} and extend them to the 
$0 < \alpha \leq 1/2$ interval. In fact, we show that there is an exact symmetry in the problem that relates
the problem with $\alpha > 1/2$ to the problem with $\alpha < 1/2$. Our method furthermore allows us to garner
additional results, such as the distribution of the time $t_m$ at which the maximum of the price is reached. 
We find that this distribution has inverse square root singularities both at $t_m=0$ and $t_m=T$ {\it for all 
values of} $\alpha$; however, the amplitude of the divergence at $t_m=0$ is stronger when $\alpha < 1/2$ and
weaker when $\alpha > 1/2$. This gives a more precise picture to the results of Shiryaev, Xu and Zhou. 
For $\alpha=1/2$, the problem is degenerate and the two peaks have exactly the same amplitude (in fact, the
distribution is symmetric under $t_m \to T-t_m$).

Finally, we do not agree with the statement that $\alpha < 1/2$ is not interesting financially. The numbers
provided in \cite{SXZ} are based on the S\&P500 index returns, and are therefore much too optimistic: first, 
there is an obvious selection bias since badly performing stocks leave the index; second, the volatility of
the index is two to three times smaller than the volatility of individual stocks, thanks to the diversification
effect. An annualized volatility above $40 \%$ is in fact not uncommon, in particular for small to medium caps --
whereas the S\&P500 only includes large caps. Fig. 1 shows the time series of the (implied) S\&P500 index 
volatility and the average stock volatility, in the period 2000-2007. With an average annual return of 
$10 \%$, an interest rate of $5\%$ and a volatility of $40 \%$ annual, the parameter $\alpha$ is found to 
be $0.3125 < 1/2$.

We hope that this short note will shed a useful light on the work of A. Shiryaev, Z. Xu and X.Y. Zhou, and
that it will convince the reader that path integral methods are extremely powerful to solve a variety of
random walk problems. We refer the reader to a short review paper by one of us \cite{BF} on this topic, see
also \cite{MRKY}.

\begin{figure}
\includegraphics[width=.9\hsize]{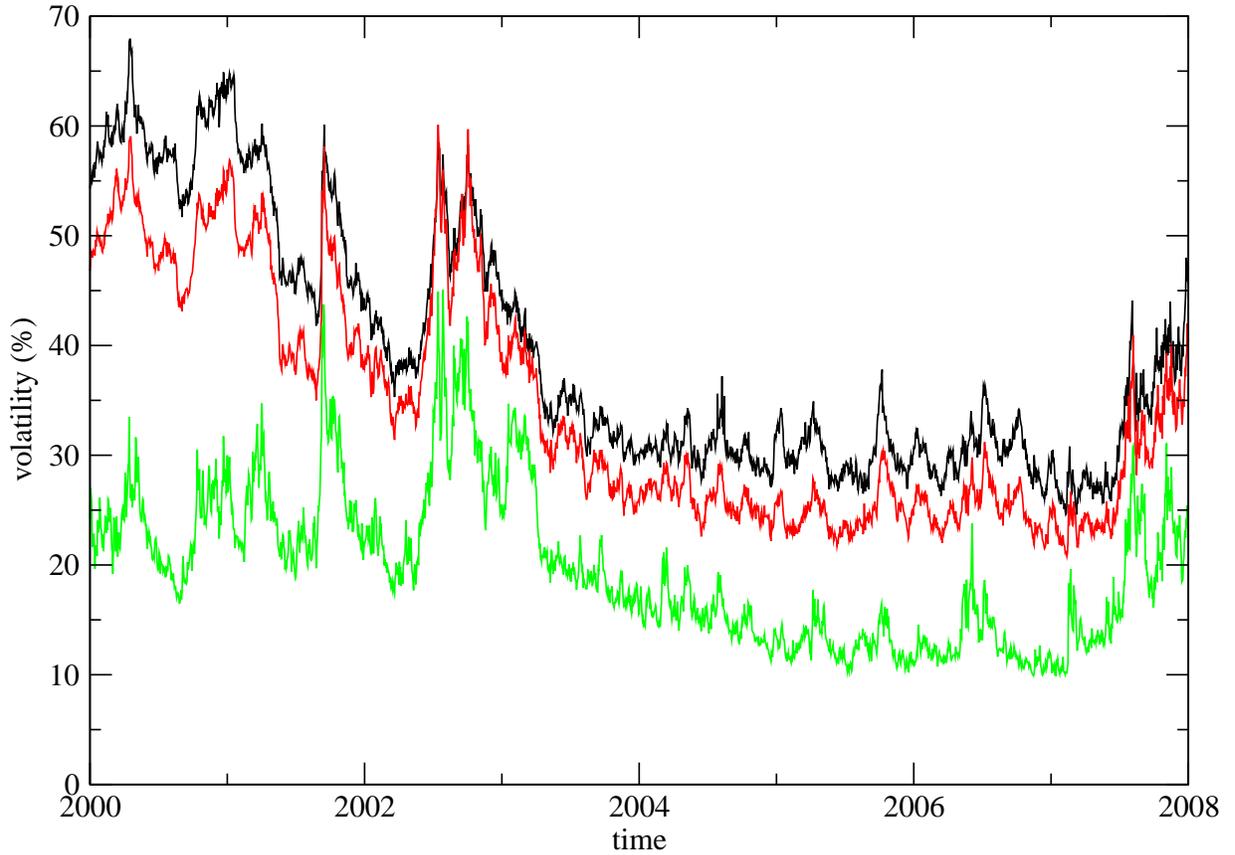}
\caption{Time series of the (implied) S\&P500 index 
volatility (the VIX, bottom green curve) and the average implied stock volatility for SPX stocks (middle red curve) and
mid-cap stocks (upper black curve), in the period 2000-2007. It is clear that stock volatilities are on average 
two to three times larger than the volatility of the index, and that values of $\sigma$ above $40 \%$ are not uncommon.} 
\end{figure} 

\section{The Set-Up}

In this section we give the set-up of the problem using physicists notations. We assume, as in \cite{SXZ}, that
the price $P_t$ of a stock follows a geometric Brownian motion:
\begin{equation}
P_t= \exp\left[\mu t + \sigma\,B_t\right],
\label{BS1}
\end{equation}
where $\mu$ is Ito corrected drift and $B_t$ the standard Brownian motion.
We will use below the notation $x(t)=\mu t + \sigma B_t$ for the drifted Brownian motion,
with by convention $x(0)=0$. In Ref. \cite{SXZ}, the authors introduce the notation $\alpha=\mu/\sigma^2+1/2$.
A `good' stock in the financial language corresponds to a positive drift $\mu>0$ (i.e., $\alpha>1/2$)
and a `bad' stock corresponds to a negative drift $\mu<0$ ($\alpha<1/2$). In terms of the real return $a$ of the stock, 
the condition $\mu >0$ translates into $a-r > \sigma^2/2$, where $a-r$ is the excess return over the risk-free rate $r$. 
Note that the process that we talk about is the real world process and not the risk-neutral one, 
which has no meaning for the question raised in Ref. \cite{SXZ}. 

Let us consider the evolution of the stock price over a fixed time interval
$0\le t \le T$. It is intuitively obvious that the maximum of a 
drifted Brownian motion 
and hence that of the stock price $P_t$ is most likely to occur at $t=T$ (for $\mu>0$)
and $t=0$ (for $\mu<0$). Thus, it obviously makes sense to sell a `good' stock $(\mu>0)$
at the end of the interval $t=T$, whereas a `bad' stock $(\mu<0)$ at the begining
of the interval $t=0$. This intuitive results are put on a more rigorous mathematical
footing in the rest of this note by (i) calculating exactly, using path integral methods, the maximal 
relative error as
defined in Ref. \cite{SXZ}, but for all values of $\mu$ and (ii) also by computing the
full probability density of the time $t_m$ at which the maximum of the price occurs for
all $\mu$.  

Let $M_T$ denote the maximum price of the stock over the
interval $[0,T]$, i.e.,
\begin{equation}
M_T = \max_{\substack{0\le t\le T}} P_t.
\label{max1}
\end{equation}
Evidently, the optimal time to sell the stock is the one where the
difference between the price of the stock and its maximal value $M_T$
is minimal. A convenient way to estimate this optimal time is to
consider the relative error at a fixed time $\tau$ where $0\le \tau\le T$
\begin{equation}
r_{\mu} (\tau,T)= 
{\rm E}\left(\frac{M_T-P_\tau}{M_T}\right)=1-{\rm E}\left(\frac{P_\tau}{M_T}\right)
\label{re1}
\end{equation}
where ${\rm E}$ denotes the expectation value over all realizations of the Brownian 
motion. Minimizing $r_{\mu}(\tau,T)$ over $0\le \tau\le T$ gives the optimal
time $\tau^*$. In other words, $\tau^*$ is the time at which the ratio
\begin{equation}
S_{\mu} (\tau,T)= 1-r_{\mu}(\tau,T)= {\rm E}\left(\frac{P_\tau}{M_T}\right)
\label{se1}
\end{equation}
is maximal. The goal is to estimate $S_{\mu}(\tau,T)$ and then maximize it
with respect to $0\le \tau \le T$. Using the 
trivial
identity
\begin{equation}
M_T= \max_{\substack{0\le \tau\le T}} P_\tau=\max_{\substack{0\le \tau\le T}} 
\left[\exp\left(x(\tau)\right)\right]= \exp\left[\max_{\substack{0\le \tau\le T}} x(\tau)\right]
\label{iden1}
\end{equation}
one can rewrite $S_\mu(\tau,T)$ in Eq. (\ref{se1}) as 
\begin{equation}
S_\mu(\tau,T)= {\rm E}\left[\exp\left(-\{\tilde M_T-x(\tau)\}\right)\right]
\label{se2}
\end{equation}
where 
\begin{equation}
\tilde M_T= \max_{\substack{0\le t\le T}} x(t)=\ln[M_T]
\label{max2}
\end{equation}
is the
maximum of the drifted Brownian motion $x(t)$ over $0\le t\le T$. Note that
throughout this paper, we will use $t$ as the running time and $\tau$ as a
fixed time.

Let us consider the random variable $y(\tau)= \tilde M_T-x(\tau)$ at a fixed time 
$\tau$ and let
$P_\mu(y,\tau)$ denote its probability density function (pdf). Once we know
$P_\mu(y,\tau)$, then from Eq. (\ref{se2}), we can evaluate 
\begin{equation}
S_\mu(\tau,T)= \int_0^{\infty} dy\, e^{-y}\, P_\mu(y,\tau).
\label{se3}
\end{equation}
To evaluate the pdf $P_\mu(y,\tau)$, we need the joint pdf of $\tilde M_T$ and
$x(\tau)$ at fixed $\tau$.

\section{Joint distribution of $\tilde M_T$ and $x(\tau)$} 

It is convenient
to compute first the cumulative probability 
\begin{equation}
F_{\mu}(x,m,\tau)= {\rm Prob}[x(\tau)=x\,\,{\rm and}\,\, \tilde M_T \le m]
\label{cum1}
\end{equation}
where the walk starts at the origin $x(0)=0$ and $\tilde M_T$ is the
global maximum of the walk in $[0,T]$. This cumulative probability 
can be computed using a path-integral approach as detailed below.

Clearly $F_\mu(x,m,\tau)$ is the probability that a drifted Brownian motion
$x(t)$ in $0\le t\le T$, starting from $x(0)=0$, reaches $x(\tau)=x$ at a 
fixed time $t=\tau$ and in addition, stays below the level $m$ for all $0\le t\le T$.
The last condition comes from the fact that if the global maximum $\tilde M_T \le m$,
the path necessarily stays below the level $m$ for all $0\le t\le T$. An example of 
such a path is seen in Fig. 2.
\begin{figure}
\includegraphics[width=.9\hsize]{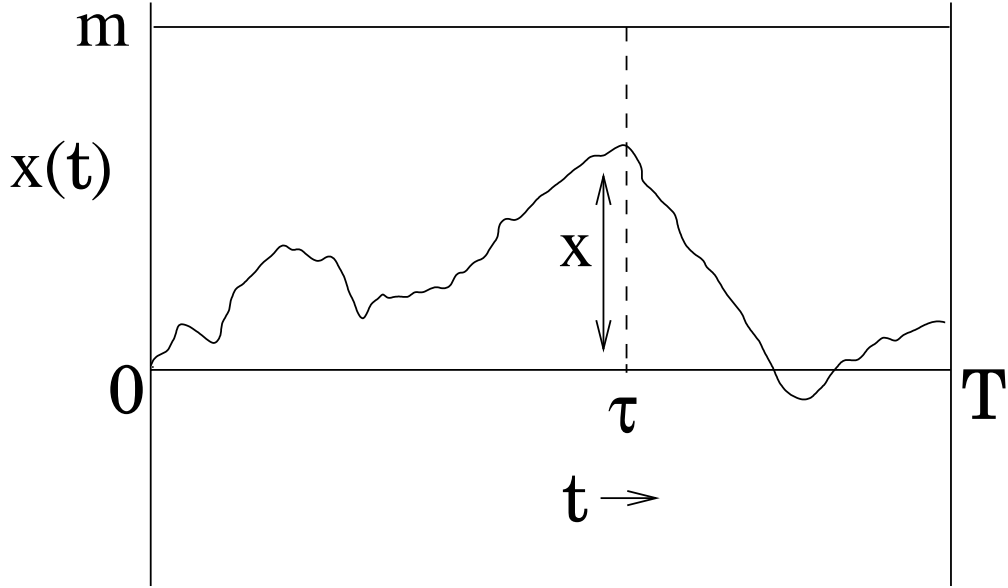}
\caption{A realization of the drifted Brownian motion $x(t)$ in $t\in [0,T]$, 
starting at $x(0)=0$, reaching $x(\tau)=x$ at $t=\tau$ and staying below the
level $m$ for all $0\le t\le T$.}
\end{figure}
To compute $F_{\mu}(x,m,\tau)$, it is convenient to consider the shifted process
$y(t)= m- x(t)$ so that the process $y(t)$ evolves, as
\begin{equation}
dy=-dx= -\mu dt -\sigma dB_t.
\label{lange2}
\end{equation}
Thus the shifted process $y(t)$ represents a Brownian motion with a drift 
$-\mu$, opposite to that of $x(t)$. In terms of the process $y(t)$, $F_{\mu}(x,m,\tau)$
is just the probability that the process $y(t)$, starting at $y(0)=m$, reaches
the point $y(\tau)= m-x$ at $t=\tau$ and {\it stays positive} in the whole interval
$0\le t\le T$. An example of such an event is shown in Fig. 3
\begin{figure}
\includegraphics[width=.9\hsize]{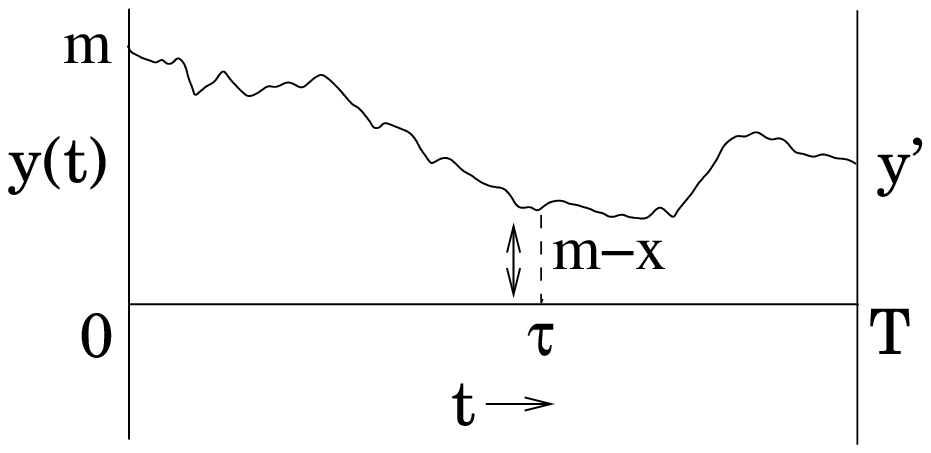}
\caption{A realization of the shifted Brownian motion $y(t)$ with
drift $-\mu$ in $t\in [0,T]$,
starting at $y(0)=m$, reaching $y(\tau)=m-x$ at $t=\tau$ and staying positive
for all $0\le t\le T$.}
\end{figure}

For the process $y(t)$ in Eq. 
(\ref{lange2}), 
let us first define the propagator $G_{-\mu}^{+}(y,y_0,t)$ that denotes the
probability that the process starting at $y_0$ at $t=0$, reaches $y$ at time $t$,
but staying {\it positive} in between, i.e., in $[0,t]$. One can then easily
express $F_{\mu}(x,m,\tau)$ in terms of this propagator as (see Fig. 3)
\begin{equation}
F_{\mu}(x,m,\tau)= G_{-\mu}^{+}(m-x,m,\tau)\,\int_0^{\infty} G_{\mu}^{+}(y',m-x,T-\tau)\, 
dy' .
\label{propa1}
\end{equation}
In writing Eq. (\ref{propa1}), we have split the interval $[0,T]$ into two
parts: $[0,\tau]$ and $[\tau,T]$. In the first interval (see Fig. 3), the
process propagates from the initial position $y(0)=m$ to $y(\tau)=m-x$
in time $\tau$ (staying positive in between), hence explaining the first factor 
$G_{-\mu}^{+}(m-x,m,\tau)$
in Eq. (\ref{propa1}). In the second interval, the process starting at
the new `initial' position $m-x$, propagates to a final position $y'$
in time $T-\tau$, staying positive in between. Also, the final position
$y'$ can be any positive number and one has to integrate over it. This
explains the second factor in Eq. (\ref{propa1}). Of course, in writing
the path decomposition in Eq. (\ref{propa1}) we have used the renewal property of a 
Brownian motion (valid due to its Markovian nature) which implies that the two intervals (left of 
$\tau$ and right of 
$\tau$) are statistically independent.

\vskip 0.3cm

{\noindent {\bf Evaluation of the propagator $G_{-\mu}^{+}(y,y_0,\tau)$:}} Using a physicist
interpretation of Eq. (\ref{lange2}), we note that the Langevin noise $\eta(t)=dB_t/dt$ is a Gaussian 
white noise with the associated measure, ${\rm Prob}\left[\{\eta(t)\}, 0\le t\le \tau \right]\propto 
\exp\left[-\frac{1}{2}\int_0^{\tau} 
\eta^2(t)dt\right]$. Substituting, $\eta(t)=({\dot y}+\mu)/\sigma$ from Eq. (\ref{lange2}),
one can express the propagator as a path integral 
\begin{equation}
G_{-\mu}^{+}(y,y_0,\tau)=\int_{y(0)=y_0}^{y(\tau)=y}{\cal D}y(t) 
\exp\left[-\frac{1}{2\sigma^2}\int_0^\tau dt \left({\dot y}+\mu\right)^2\right]\, 
\left[\prod_{t=0}^\tau 
\theta(y(t))\right] 
\label{green1}
\end{equation}
where $\left[\prod_{t=0}^\tau
\theta(y(t))\right]$ is an indicator function that enforces the path to stay positive in
the interval $t\in [0,\tau]$. The rhs of Eq. (\ref{green1}) can be rearranged (by expanding
the square $({\dot y}+\mu)^2$ and performing the time integral) as 
\begin{equation}
G_{-\mu}^{+}(y,y_0,\tau)= \exp\left[-\frac{\mu^2\tau}{2\sigma^2}-\frac{\mu}{\sigma^2}(y-y_0)\right]\, 
G_0^{+}(y,y_0,\tau)
\label{green2}
\end{equation}
where $G_0^{+}(y,y_0,\tau)$ is the propagator associated with the driftless $(\mu=0)$ Brownian motion
\begin{equation}
G_0^{+}(y,y_0,\tau)= \int_{y(0)=y_0}^{y(\tau)=y}{\cal D}y(t)
\exp\left[-\frac{1}{2\sigma^2}\int_0^\tau dt \, {\dot y}^2\right]\, \left[\prod_{t=0}^\tau
\theta(y(t))\right].
\label{greenfree}
\end{equation}
This propagator, which denotes the probability that a driftless Brownian motion
propagates from $y_0$ to $y$ in time $\tau$ without crossing the origin in between,
can be evaluated 
very simply by the standard method of images~\cite{Feller,Redner} or
alternatively by the path integral method~\cite{BF} and has the
well known expression
\begin{equation}
G_0^{+}(y,y_0,\tau)= \frac{1}{\sqrt{2\pi \sigma^2 
\tau}}\,\left[\exp\left(-\frac{(y-y_0)^2}{2\sigma^2\tau}\right)
-\exp\left(-\frac{(y+y_0)^2}{2\sigma^2\tau}\right)\right].
\label{greenfree1}
\end{equation}
Substituting this in Eq. (\ref{green2}), one then has the required propagator.
Using this explicit expression for $G_{-\mu}^{+}(y,y_0,\tau)$ one can also
easily evaluate the following integral
\begin{equation}
\int_0^{\infty} G_{-\mu}^{+}(y,y_0,\tau)\, dy = \frac{1}{2}\,
\left[ {\rm erfc}\left(-\frac{y_0-\mu \tau}{\sqrt{2\sigma^2 \tau}}\right)- 
\exp\left(\frac{2\mu\,y_0}{\sigma^2}\right)\,{\rm erfc}\left(
\frac{y_0+\mu
\tau}{\sqrt{2\sigma^2 \tau}}\right)\right]
\label{righthalf}
\end{equation}
where ${\rm erfc}(x)= \frac{2}{\sqrt{\pi}}\int_x^{\infty} e^{-u^2}\,du$ is the complementary
error function.
Assembling these results in Eq. (\ref{propa1}) gives us an explicit expression for
the cumulative probability
\begin{equation}
F_{\mu}(x,m,\tau)=\frac{e^{-\frac{\mu^2\tau}{2\sigma^2}+\frac{\mu x}{\sigma^2}}}{2\sqrt{2\pi 
\sigma^2\tau}}\, 
\left[e^{-\frac{x^2}{2\sigma^2\tau}}
-e^{-\frac{(2m-x)^2}{2\sigma^2\tau}}\right]\,
\left[ {\rm erfc}\left(-\frac{m-x-\mu (T-\tau)}{\sqrt{2\sigma^2 (T-\tau)}}\right)-
e^{\frac{2\mu\,(m-x)}{\sigma^2}}\,{\rm erfc}\left(
\frac{m-x+\mu
(T-\tau)}{\sqrt{2\sigma^2 (T-\tau)}}\right)\right].
\label{fxm}
\end{equation}
The joint pdf $Q_{\mu}(x,m)$ of $x(\tau)=x$ at fixed $\tau$ and $\tilde M_T=m$ can then 
be 
obtained by taking the derivative of $F_{\mu}(x,m,\tau)$ with respect to $m$, i.e.,
\begin{equation}
Q_{\mu}(x(\tau)=x, \tilde M_T=m)= \frac{\partial F_{\mu}(x,m,\tau)}{\partial m}.
\label{jpdfQ}
\end{equation}  

\section{Evaluation of the relative error $r_{\mu}(\tau,T)$}

Having obtained the joint pdf $Q_{\mu}(x(\tau)=x, \tilde M_T=m)$ in Eqs. (\ref{jpdfQ})
and (\ref{fxm}), we can easily find the pdf $P_\mu(y,\tau)$ of the variable $y=\tilde M_T-x$
\begin{eqnarray}
P_{\mu}(y,\tau)&=& \int Q_{\mu}(x, m)\, \delta\left(y-(m-x)\right)\, dx\, dm \nonumber \\
&=& \int_0^{\infty} Q_{\mu}(m-y,m)\, dm.
\label{pdfy1}
\end{eqnarray}
The above integral can be performed exactly (we skip the details here). One obtains
the following expression 
\begin{equation}
P_{\mu}(y,\tau)= \frac{1}{\sqrt{2\pi \sigma^2\tau}}\,f_{\mu}(y,\tau)\,g_{\mu}(y,T-\tau)
+\frac{1}{\sqrt{2\pi \sigma^2(T-\tau)}}\,f_{-\mu}(y,T-\tau)\,g_{-\mu}(y,\tau)
\label{pdfy2}
\end{equation}
where
\begin{eqnarray}
f_{\mu}(y,\tau)&=& 
\exp\left(-\frac{(y+\mu\tau)^2}{2\sigma^2\tau}\right)+\frac{\mu}{\sigma}
\sqrt{\frac{\pi \tau}{2}}\,e^{-2\mu y/\sigma^2}\,{\rm erfc}\left(\frac{y-\mu 
\tau}{\sqrt{2\sigma^2\tau}}\right) \label{f1}\\
g_{\mu}(y,\tau)&=& \left[ {\rm erfc}\left(-\frac{y-\mu \tau}{\sqrt{2\sigma^2 \tau}}\right)-
e^{2\mu\,y/\sigma^2}\,{\rm erfc}\left(
\frac{y+\mu\tau}{\sqrt{2\sigma^2 \tau}}\right)\right]
\label{g1}
\end{eqnarray}  
Note from the explicit expression of $P_\mu(y,\tau)$ the following symmetry
\begin{equation}
P_{\mu}(y,\tau)= P_{-\mu}(y, T-\tau)
\label{symm1}
\end{equation}
which has the simple physical meaning of time-reversal symmetry, i.e., when the
process propagates in the reverse time direction, one gets the same measure
provided one also reverses the sign of the drift $\mu$.

Having obtained the pdf $P_\mu(y,\tau)$, one can then evaluate the relative 
error $r_\mu(\tau,T)=1-S_\mu(\tau,T)$ where
\begin{equation}
S_{\mu}(\tau,T)= \int_0^{\infty} dy\, e^{-y} P_{\mu}(y,\tau)
\label{sm1}
\end{equation}
Evidently $S_{\mu}(\tau,T)$ also has the same time-reversal symmetry namely
\begin{equation}
S_{\mu}(\tau,T)= S_{-\mu}(T-\tau, T)
\label{symm2}
\end{equation}
While it is difficult to do the integral in Eq. (\ref{sm1}) analytically, 
one can easily evaluate it 
using Mathematica. Besides, the general feature of $S_{\mu}(\tau,T)$ as a function
of $\tau$ can be inferred by just studying the asymptotic properties of the integral in Eq. 
(\ref{sm1}) in the
limit $\tau\to 0$ and $\tau \to T$. In Fig. 4, we 
show a 
plot of $S_{\mu}(\tau,T)$ for three different values
of $\mu=0.1$, $\mu=0$ and $\mu=-0.1$ upon setting $T=1$ and $\sigma=1$.
\begin{figure}
\includegraphics[width=.9\hsize]{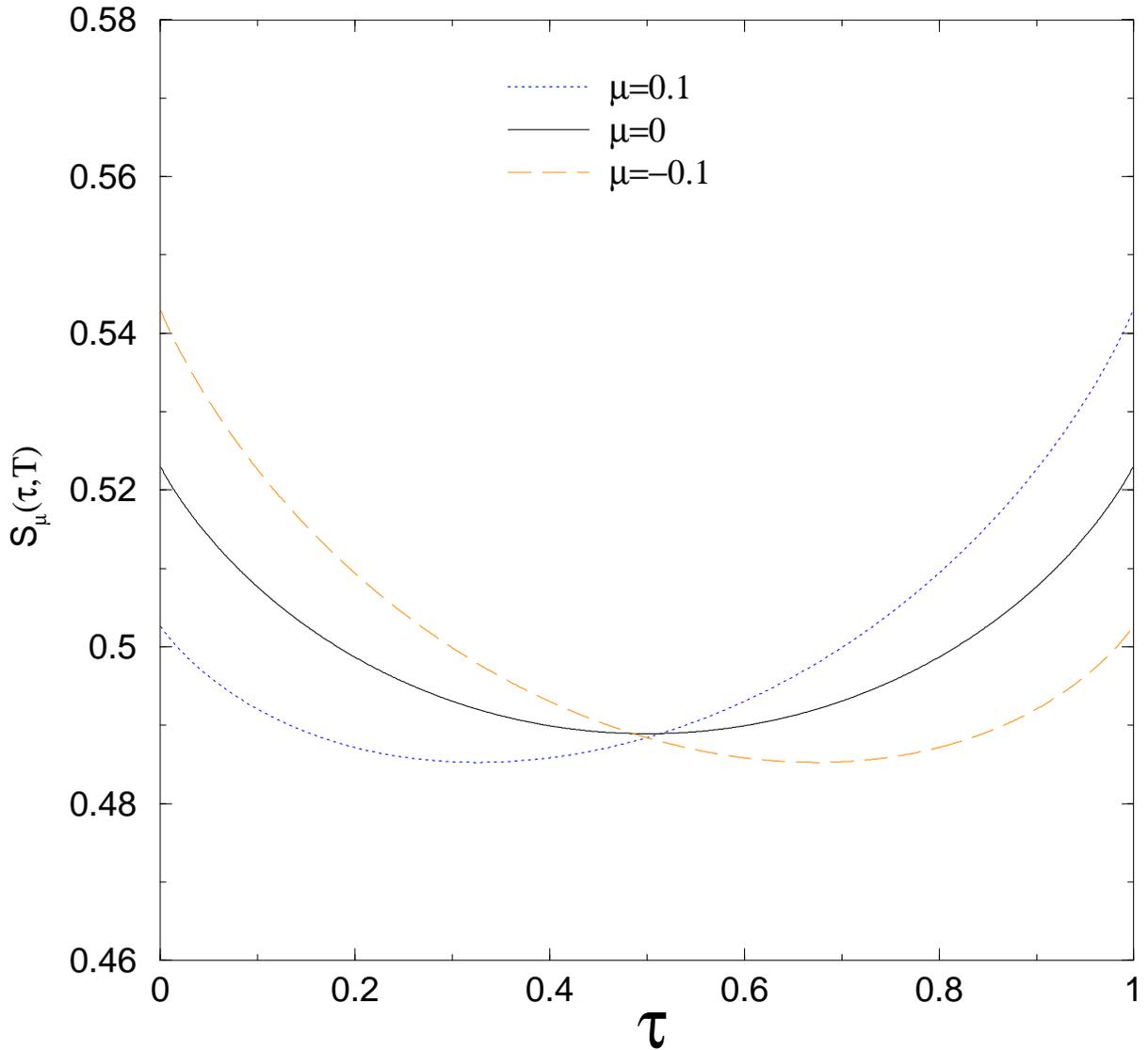}
\caption{Plots of $S_{\mu}(\tau, T)$ vs. $\tau$ obtained from Eq. (\ref{sm1}) for three different 
values of the drift
$\mu=0.1$, $\mu=0$ and $\mu=-0.1$. We have set $T=1$ and $\sigma=1$. The symmetry 
$S_\mu(\tau,T)=S_{-\mu}(T-\tau,T)$ is
evident.} 
\end{figure} 

\vskip 0.3cm

{\noindent {\bf Optimal Time $\tau^*$:} To find the optimal time $\tau^*$ we need to minimize
$r_\mu(\tau,T)$, i.e., maximize $S_\mu(\tau,T)$ with respect to $\tau\in [0,T]$. It is 
evident from Fig. 4 and also from the expression of $S_{\mu}(\tau,T)$ that for all values of $\mu$,
$S_{\mu}(\tau,T)$ has two local maxima at the endpoints of the interval $[0,T]$, i.e., respectively 
at 
$\tau=0$ and $\tau=T$. However, for $\mu>0$, the maximum at $\tau=T$ has a larger value implying
that for $\mu>0$, $\tau^*=T$. By the symmetry manifest in $S_{\mu}(\tau, T)=S_{-\mu}(T-\tau,T)$ it 
follows that for $\mu<0$, the maximum at $\tau=0$ has a higher value implying $\tau^*=0$ for $\mu<0$.
Exactly at $\mu=0$, both local maxima at $\tau=0$ and $\tau=T$ have the same value 
($S_0(\tau,T)$ is completely symmetric around the midpoint $\tau=T/2$) implying that
for $\mu=0$, both $\tau^*=0$ and $\tau^*=T$ are optimal.

The optimal value $S_{\mu}(\tau^*,T)$ is actually easier to evaluate since for $\tau=0$ or 
$\tau=T$ (at the end-points), the integral in Eq. (\ref{sm1}) can be carried out explicitly. 
Omitting details of this integration, we get the following expression for the optimal relative error 
for all $\mu$
\begin{eqnarray}
r(\tau^*,T)&=& 1-S_{\mu}(\tau^*,T) \nonumber \\
&=& 1- \frac{|\mu|}{2|\mu|+\sigma^2}\,{\rm erfc}\left(-\frac{|\mu|}{\sigma}\sqrt{\frac{T}{2}}\right)-
\frac{\sigma^2+|\mu|}{\sigma^2+2|\mu|} \, \exp\left[-\left(|\mu|+\frac{\sigma^2}{2}\right)\,T\right] 
\,{\rm 
erfc}\left(\left(\frac{|\mu|}{\sigma^2}+1\right)\sqrt{\frac{\sigma^2T}{2}}\right).
\label{opterr}
\end{eqnarray}
Note that the optimal relative error is evidently a symmetric function of $\mu$ as manifest
in the above result.

In the preceeding paper~\cite{SXZ}, Shiryaev, Xu and Zhou also obtained an expression of 
the optimal relative error $r(\tau^*,T)$ by a completely different method.
Their notations are slightly different from above. In their notation, $\mu=(\alpha-1/2)\sigma^2$
and also their result for $r(\tau^*,T)$ is in terms of the probability distribution of a Gaussian 
random variable
with zero mean and unit variance, $\Phi(x)= \frac{1}{\sqrt{2\pi}}\int_{-\infty}^{x} e^{-u^2/2} du$
which is related to the complimentary error function via
\begin{equation}
\Phi(x)= \frac{1}{2}\, {\rm erfc}\left(-\frac{x}{\sqrt{2}}\right).
\label{norm1}
\end{equation}
However, their method allows them to obtain an explicit expression for the optimal relative error
only in the range $\alpha\ge 1/2$ (i.e., $\mu\ge 0$) and $\alpha\le 0$ (i.e., $\mu\le -\sigma^2/2$).
In these ranges, their expressions for the optimal relative error (Eqs. (9) and (11) in \cite{SXZ})
reduce precisely to our compact result in Eq. (\ref{opterr}), upon identifying
$\mu=(\alpha-1/2)\sigma^2$ and $\Phi(x)$ as in Eq. (\ref{norm1}). However,
they do not have any result in the range $0<\alpha<1/2$, i.e., for $-\sigma^2/2<\mu<0$. In contrast,
our result in Eq. (\ref{opterr}) is valid for all $\mu$ (and hence for all $\alpha$) and
is therefore more general. In addition, their method somehow does not detect the symmetry
of $r(\tau^*,T)$ under $\mu\to -\mu$ which is manifest in our path integral approach.

\section{The exact distribution of the time $t_m$ of the occurrence of the maximum for a Brownian 
motion
with drift $\mu$}

Minimizing the relative error $r(\tau,T)$ with respect to $\tau$ is one way of estimating
the optimal time $\tau^*$ at which one should sell a stock over a fixed investment time 
horizon $T$, as explained above. Another alternative and direct measure would be to
first derive the probability density $p(t_m,T)$ of the time $t_m$ at which the maximum $M_T$ of a
stock price over $[0,T]$ actually occurs. This density $p(t_m,T)$ will typically have
a peak (or more peaks). The value of $t_m=t^*$ at which the strongest peak of $p(t_m,T)$
occurs can then be taken as an alternative measure for the optimal time to sell a stock, since the
maximum of the price is most likely to occur at $t_m=t^*$. 

In this section, we compute exactly the density $p_\mu(t_m,t)$ of $t_m$ for a Brownian motion 
$x(t)$ with 
drift $\mu$.
Since the stock price $P_t=\exp[x(t)]$ is just the exponential of $x(t)$ under the Black-Scholes 
scenario,
the maximum $M_T$ of the stock price $P_t$ occurs exactly at the same time $t_m$ where
$x(t)$ itself achieves its maximum.  
For the case $\mu=0$, the density $p_0(t_m,T)$ was computed by L\'evy~\cite{Levy} and is given 
by the derivative of an arcsine form, i.e.,
\begin{equation}
p_0(t_m,T)= \frac{1}{\pi}\, \frac{1}{\sqrt{t_m(T-t_m)}}; \quad\, 0\le t_m \le T
\label{levy1}
\end{equation}
Recently, using an appropriate path integral method, the density of $t_m$ was computed exactly for a 
Brownian motion up to
its first-passage time ~\cite{RM} and also for a variety of constrained Brownian motions
such as Brownian excursions, Brownian bridges, Brownian meanders  etc.~\cite{MRKY}. Here we adapt 
this
path integral method to compute the density $p_{\mu}(t_m,T)$ for a Brownian motion  
with arbitrary drift $\mu$.

To compute the density $p_{\mu}(t_m,T)$ the strategy would be to first compute
the joint density of $t_m$ as well as the maximum $\tilde M_T=m$ itself, i.e., $p_{\mu}(t_m,m,T)$
and then integrate over $m$ to obtain the marginal density, $p_\mu(t_m,T)= \int_0^{\infty} 
p_\mu(t_m,m,T)\,dm$. The joint density $p_\mu(t_m,m,t)$ is the proportional to the
sum of the statistical weights of all paths that start at the origin $x(0)=0$, reaches
the value $x(t_m)=m$ for the {\it first-time} at $t=t_m$ and then stays below the level $m$
at all subsequent times up to $T$,i.e., in the interval $[t_m,T]$. To enforce the conditions
that $x(t)<m$ in the two intervals $t\in [0,t_m]$ and $t\in [t_m,T]$ and that 
exactly at $t_m$ the
path reaches 
$x(t_m)=m$, poses a problem for a continuous-time
Brownian motion. This is because if a Brownian motion crosses a level $m$ at a given time $t_m$
then it must cross and re-cross the same level $m$ an infinite number of times in the
vicinity of $t=t_m$. Hence it is impossible to enforce the above constraints simultaneously for
a continuous-time Brownian motion. Note that for lattice random walks this
does not pose any problem. To get around this 
difficulty with the continuous-time Brownian motion,
one introduces a small cut-off $\epsilon$~\cite{RM,MRKY}, i.e., one assumes that the path, starting 
at $x(0)=0$
reaches the level $m-\epsilon$ at time $t_m$, staying below $m$ for all $0\le t< t_m$ 
and then starting at $m-\epsilon$ at $t=t_m$ stays below the level $m$ for all $t_m<t\le T$ (see
Fig. 5 for such a realization). Finally one takes the limit 
$\epsilon\to 0$ at the
end of the calculation.
\begin{figure}
\includegraphics[width=.9\hsize]{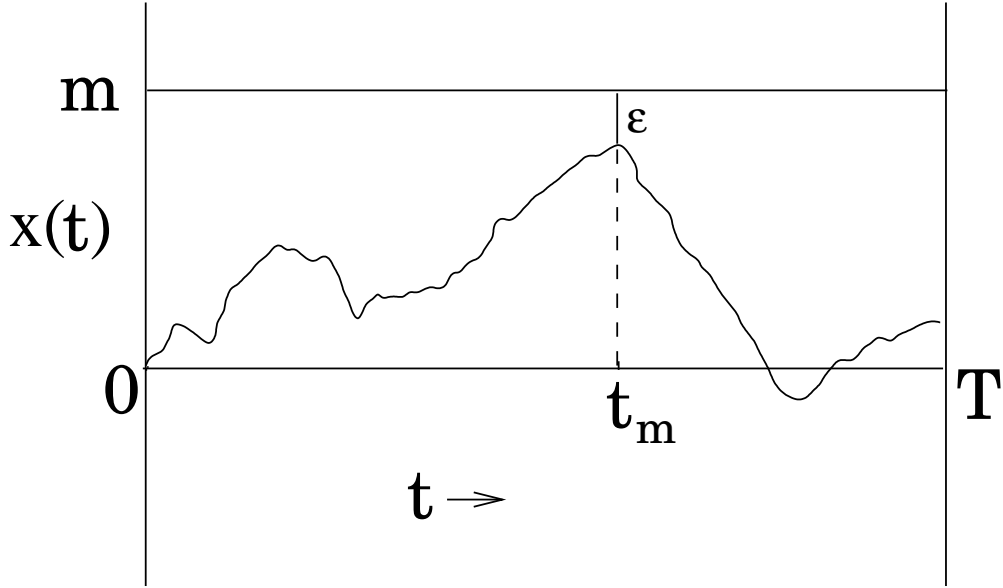}
\caption{A realization of the drifted Brownian motion $x(t)$ in $t\in [0,T]$,
starting at $x(0)=0$, reaching $x(t_m)=m-\epsilon$ at $t=t_m$ and staying below the
level $m$ for all $0\le t\le T$.}
\end{figure}

Comparing Figs. (2) and (5), it is clear that the paths that contribute to the joint probability
density $p_{\mu}(t_m,m,T|\epsilon)$ are identical to those that contribute to $F_{\mu}(x,m,\tau)$ 
with the replacements $x=m-\epsilon$ and $\tau=t_m$ in Eq. (\ref{fxm}), i.e., 
$p_{\mu}(t_m,m,T|\epsilon)\propto F_{\mu}(m-\epsilon,m,t_m)$. Substituting $x=m-\epsilon$
and $\tau=t_m$ in Eq. (\ref{fxm}) and taking the $\epsilon\to 0$ limit
we find, to leading order in $\epsilon$,
\begin{equation}
p_{\mu}(t_m,m,T|\epsilon)\xrightarrow[\epsilon\to 0]{} A\, \epsilon^2\, \frac{m 
e^{-(m-\mu t_m)^2/{2\sigma^2\, t_m}}}{\sqrt{2\pi\, \sigma^6\, t_m^3}}\,\left[\frac{2}{\sqrt{2\pi 
\sigma^2 
(T-t_m)}}\,e^{-\mu^2(T-t_m)/{2\sigma^2}}-\frac{\mu}{\sigma^2}\, {\rm 
erfc}\left(\frac{\mu}{\sigma}\sqrt{\frac{T-t_m}{2}}\right)\right]
\label{tmdist1}
\end{equation}
where the constant of proportionality $A$, which is function of $\epsilon$, is determined
from the normalization, $\int_0^T dt_m \int_0^{\infty} dm\,  p_{\mu}(t_m,m,T|\epsilon\to 0)=1$.
This fixes $A= \sigma^2/{\epsilon^2}$. Integrating $p_{\mu}(t_m,m,T)$ (now the cut-off
$\epsilon$ has been set to $0$) over $m$ finally gives the marginal density $p_{\mu}(t_m,T)$
in a closed form
\begin{equation}
p_{\mu}(t_m,T)= \frac{1}{\pi \sqrt{t_m(T-t_m)}}\, h(t_m,\mu)\, h(T-t_m,-\mu)
\label{tmdens1}
\end{equation}
where
\begin{equation}
h(t_m,\mu)= \exp\left(-\frac{\mu^2t_m}{2\sigma^2}\right)+\frac{\mu}{\sigma}\,\sqrt{\frac{\pi 
t_m}{2}}\,{\rm erfc}\left(-\frac{\mu}{\sigma}\sqrt{\frac{t_m}{2}}\right).
\label{htm1}
\end{equation}

The density $p_{\mu}(t_m,T)$ given in Eqs. (\ref{tmdens1}) and (\ref{htm1}) is the main result of 
this section.
Evidently, for $\mu=0$, one recovers from this the well known arcsine result of L\'evy in Eq. 
(\ref{levy1}).
Note that the density $p_{\mu}(t_m,T)$ also has a symmetry similar to that in Eq. (\ref{symm2})
namely
\begin{equation}
p_{\mu}(t_m,T)= p_{-\mu}(T-t_m,T).
\label{psymm}
\end{equation}
This symmetry is also evident in Fig. 6 where we plot the density $p_{\mu}(t_m,T)$ in Eq.
(\ref{tmdens1}) for $\mu=1$, $\mu=0$ and $\mu=-1$ upon setting $T=1$ and $\sigma=1$.
\begin{figure}
\includegraphics[width=.9\hsize]{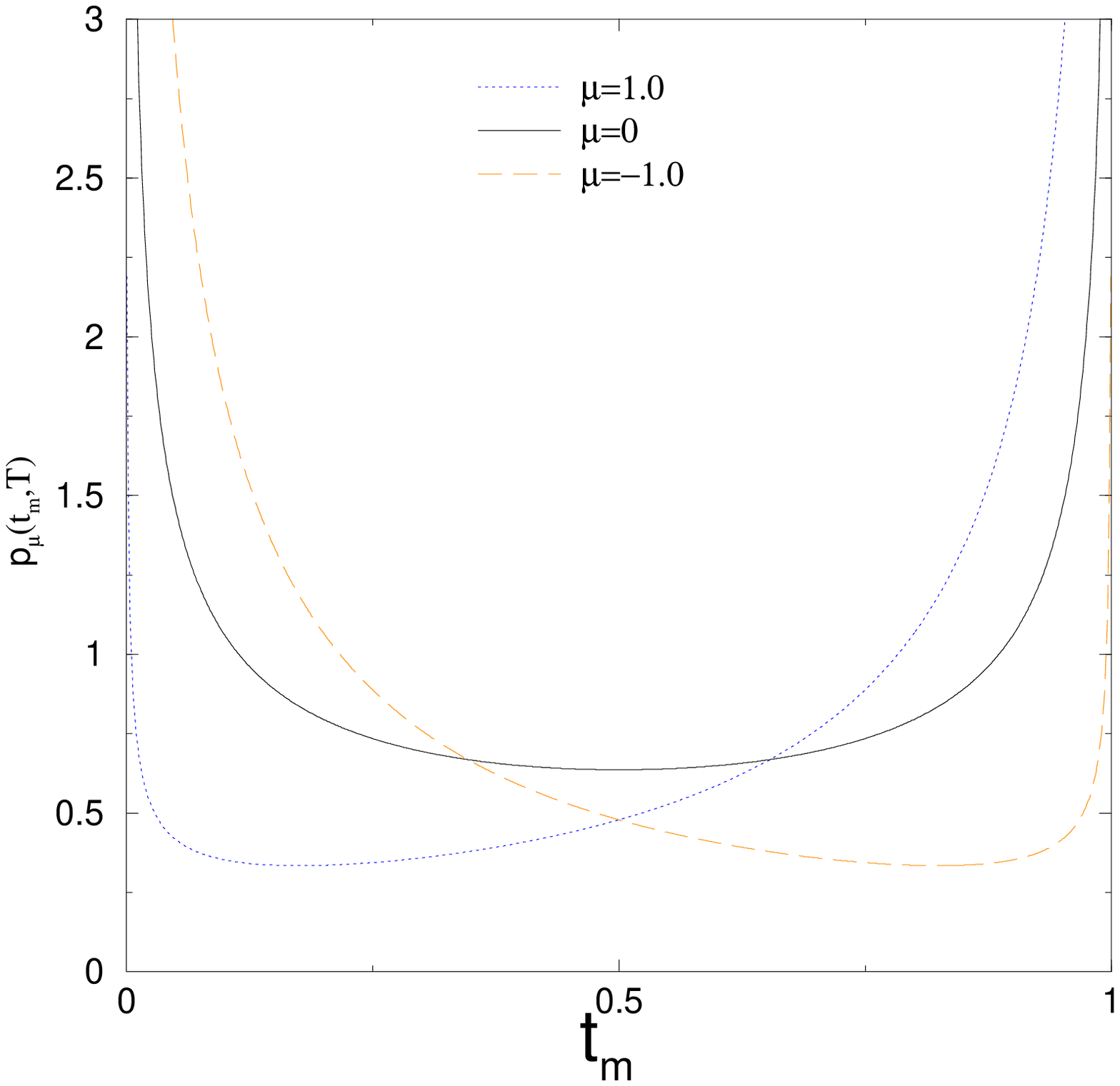}
\caption{Plots of $p_{\mu}(t_m, T)$ vs. $t_m$ for three different values of the drift
$\mu=1.0$, $\mu=0$ and $\mu=-1.0$. We have set $T=1$ and $\sigma=1$. The symmetry
$p_\mu(t_m,T)=p_{-\mu}(T-t_m,T)$ is
evident.}
\end{figure}

We note from Eq. (\ref{tmdens1}) as well as from Fig. 6 that for all values of $\mu$, the density 
$p_{\mu}(t_m,T)$ has two peaks (actually has square root divergences) at the two end points $t_m=0$ 
and $t_m=T$,
\begin{eqnarray}
p_{\mu}(t_m\to 0, T) &\approx& \frac{A_\mu(T)}{\sqrt{t_m}} \label{tml} \\
p_{\mu}(t_m\to T, T) &\approx & \frac{A_{-\mu}(T)}{\sqrt{T-t_m}} \label{tmr}
\end{eqnarray}
where the amplitude
\begin{equation}
A_{\mu}(T) = \frac{1}{\pi 
\sqrt{T}}\,\left[\exp\left(-\frac{\mu^2T}{2\sigma^2}\right)-\frac{\mu}{\sigma}\,\sqrt{\frac{\pi 
T}{2}}\,{\rm erfc}\left(\frac{\mu}{\sigma}\sqrt{\frac{T}{2}}\right)\right].
\label{ampl1}
\end{equation}
However, for $\mu>0$, the divergence at $t_m=T$ is stronger than that at $t_m=0$ since $A_{-\mu}(T)> 
A_{\mu}(T)$. On the other hand, for $\mu<0$, the opposite is true. At $\mu=0$, both ends have
the same divergences as the density is completely symmetric around $t_m=T/2$. Thus, we conclude
that the maximum of the Brownian motion with drift $\mu$ is most likely to occur at $t_m=T$
for $\mu>0$, at $t_m=0$ for $\mu<0$, and for $\mu=0$ both $t_m=0$ and $t_m=T$ are equally likely.
This then leads us to identify the optimal time $\tau^*$ to sell the stock (within the black-Scholes 
economy model) to be $\tau^*=T$ for $\mu>0$, $\tau^*=0$ for $\mu<0$, and $\tau^*=0, T$ (equally 
likely) for $\mu=0$. Thus, based on the analysis of the density 
$p_{\mu}(t_m,T)$ we draw the same 
conclusion as was obtained from the optimization of the relative error in the previous sections.

\end{document}